\documentclass[letter,notitlepage,12pt,showkeys]{revtex4-1}

\usepackage[]{graphicx}
\usepackage{amsmath}
\usepackage{cmap}

\begin{document}

\title{The nature of pulses delayed by 5~$\mu$s in scintillation detectors from showers with the energy above $10^{17}$~eV}

\author{S.~P.~Knurenko}
\author{A.~Sabourov}
\affiliation{Yu.~G.~Shafer Institute of cosmophysical research and aeronomy}
\email{s.p.knurenko@ikfia.ysn.ru}

\begin{abstract}
  Here we consider EAS events with energy above $10^{17}$~eV with recorded pulses delayed by $\tau \ge 5~\mu$s in scintillation detectors with different thresholds: $10$, $5$ and $1.8$~MeV. In order to identify pulses from electrons, muons and neutrons, experimental data were compared to computational results performed within the framework of QGSJET01d model. Preliminary, one may speculate of registration of low-energy electrons arisen from moderation of neutrons in a detector or a medium surrounding a detector or in the snow cover and frozen crust (albedo particles). The fact that such pulses were registered mostly in low-threshold detectors confirms this hypothesis.
\end{abstract}

\keywords{extensive air showers, time-related measurements}

\maketitle

\section{Introduction}

Measurements of time-related parameters of various detectors response have been provided by many experiments. In the field of ultra-high energy they are large arrays, such as Volcano Ranch, Haverah Park, Yakutsk and AGASA. In recent years similar measurements are provided at giant array Pierre Auger Observatory (PAO) with the use of water-filled Cherenkov tanks. All these surveys share one peculiarity: in some large extensive air showers (EAS) there are registered pulses from particles delayed relatively to registration of the initial particle by $\tau_{\text{d}} >> 5~\mu$s, while according to model calculations all main particles of EAS disk must arrive compactly within $\sim 1 - 5~\mu$s. In works~\cite{bib1, bib2} we have shown that in showers with $E_{0} \ge 10^{19}$~eV under certain circumstances in the time-base of scintillation detector response there pulses are observed delayed by time exceeding $5~\mu$s. According to~\cite{bib3}, one can speak of observing a nucleon component at the observation level at large distances from shower axis, namely~--- neutrons. In order to interpret the obtained results we have performed calculations of the neutrons arrival time distribution in showers with $E_{0} \ge 10^{18}$~eV for zenith angles $\theta$ from $0^{\circ}$ to $60^{\circ}$. We used CORSIKA code (version 6.900) compiled with QGSJET01d model~\cite{bib4}.

\section{The setup for time-related measurements}

A brief description of the setup for registering EAS particles with detectors of different types is given in~\cite{bib1, bib2}. The setup includes three measurement points separated bu $300-500$~m from each other. Synchronization is provided via GPS system. The ADC system, which records the pulses, is controlled by triggering signals from the main array and the small setup. After the registering program is started, the ADC continuously converts signals from detectors and cyclically records counts into the buffer memory called ``pre-history''. I.e. the ``pre-history'' stores the most recent signals. The interval between closest counts amounts to $4-10$~ns. When triggering signal arrives, it is passed to all ADC boards, and the other area of buffer memory if filled, called the ``history''. The ``history'' stores codes of ``coloring'' of the master signal~--- was it from the main array or from the small Cherenkov setup~--- and signals from detectors operated from calibrating LEDs.

\begin{figure}[htb]
  \centering
  \includegraphics[width=0.75\textwidth,clip]{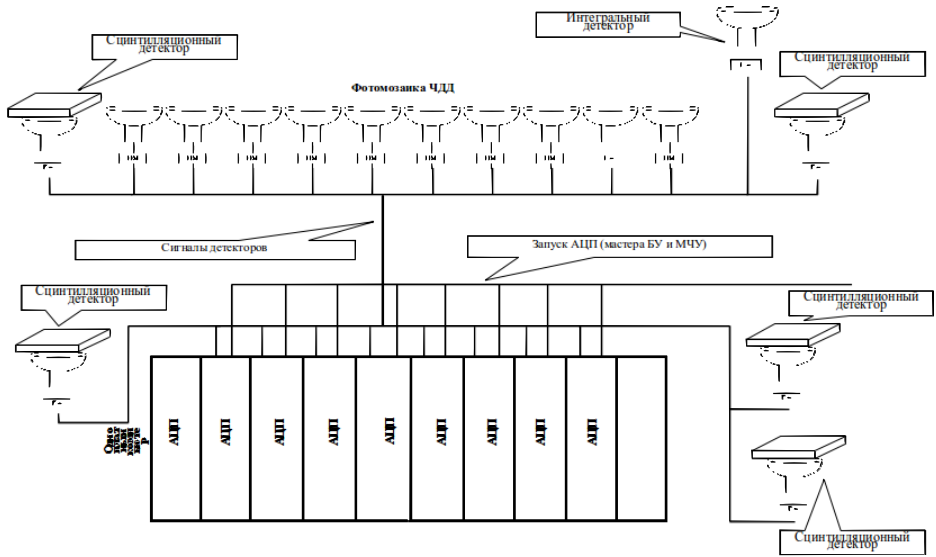}
  \caption{A diagram showing one of the setups recording the time-base of the signals from EAS particles in scintillation, Cherenkov and muon detectors.}
  \label{fig1}
\end{figure}

\section{Measurement of the temporal structure of EAS disk with the use of scintillation detectors with different thresholds}

The analysis of signals implies treatment of single detectors and their combinations. The point is that detectors have different thresholds, $10.5$, $1.8$ and $0.2$~MeV, and from comparing their signals it would be possible to estimate the energy of arriving particles and their type. The time distribution of pulses on the time-base allows to measure not only the number of registered particles but also the curvature of a shower disk and its thickness if data were recorded at different distances from a shower axis. In this way, besides integral characteristics derived from lateral distributions of various shower components it is possible to distinguish showers by the structure of detector response. For example, there is a pronounced difference in the shape of pulses recorded at different distances from the axis, especially in inclined showers where this difference arises from the separation into over-axial and under-axial particles. The difference in pulse shapes is also observed in readings of surface and underground muon detectors. It was noted, that in strongly inclined showers there are pulses delayed by significant time.

\subsection{Distribution of delayed pulses}

\begin{figure}
  \centering
  \includegraphics[width=0.75\textwidth, clip]{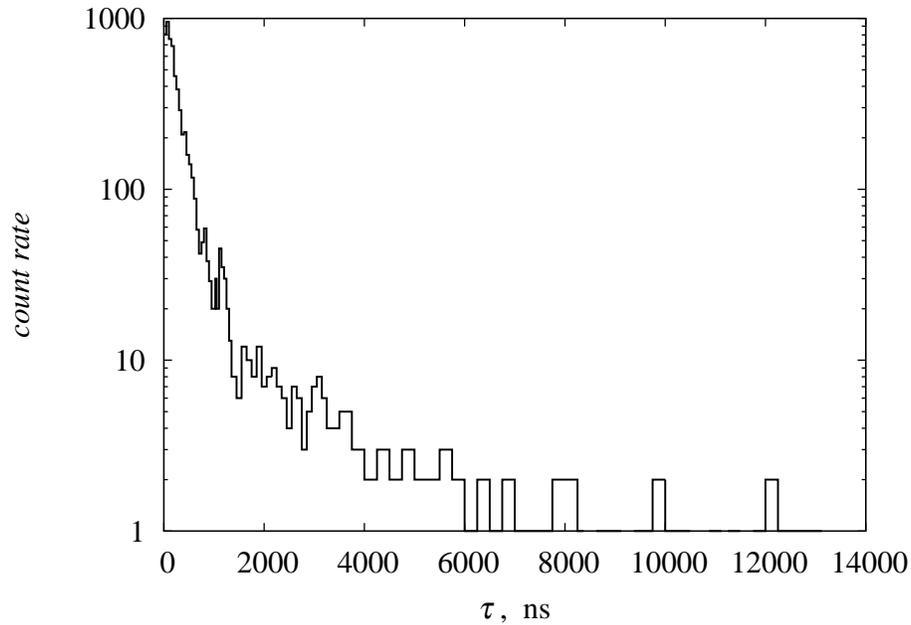}
  \caption{Temporal distribution of delayed EAS particles.}
  \label{fig2}
\end{figure}

On the Fig.~\ref{fig2} a temporal distribution is shown of pulses recorded in showers with the energy above $10^{17}$~eV. More than 6000 showers were included in the sample, with zenith angle between $0^{\circ}$ and $70^{\circ}$. It is seen that the main fraction of particles arrives within $2-3~\mu$s and only a small number of showers (mainly inclined with energies above $5 \times 10^{18}$~eV) has pulses delayed by more than $3-5~\mu$s. It is not entirely clear what sort of particles the are.

For this work we have performed calculation of lateral distribution functions (LDF) for hadrons, muons, electrons and gamma-photons in showers with $E_{0} \ge 10^{18}$~eV with the use of CORSIKA code (v.6900) and QGSJET01d model~\cite{bib4}.

\begin{figure}
  \centering
  \includegraphics[width=0.75\textwidth, clip]{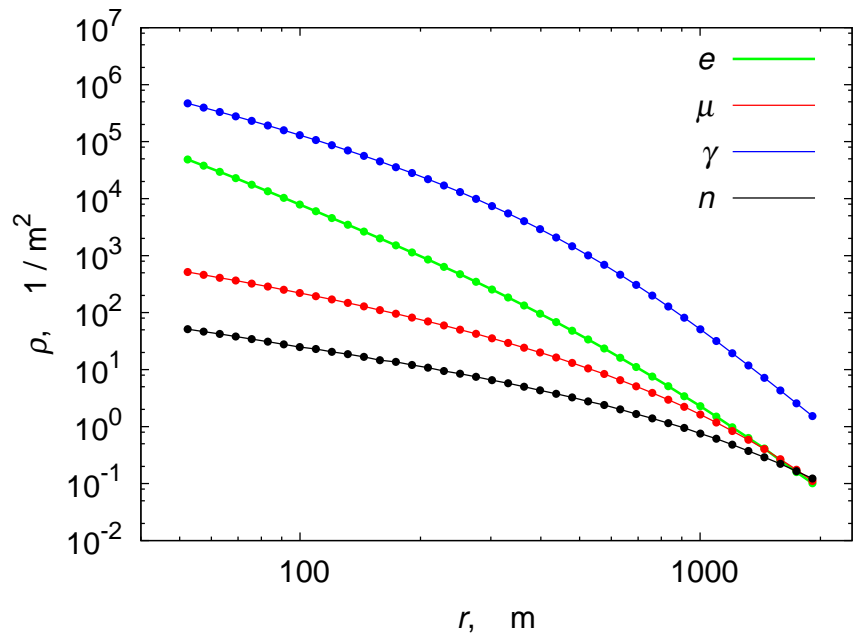}
  \caption{Lateral distributions of hadrons, muons, electrons and gamma-photons in EAS.}
  \label{fig3}
\end{figure}

\begin{figure}
  \centering
  \includegraphics[width=0.75\textwidth, clip]{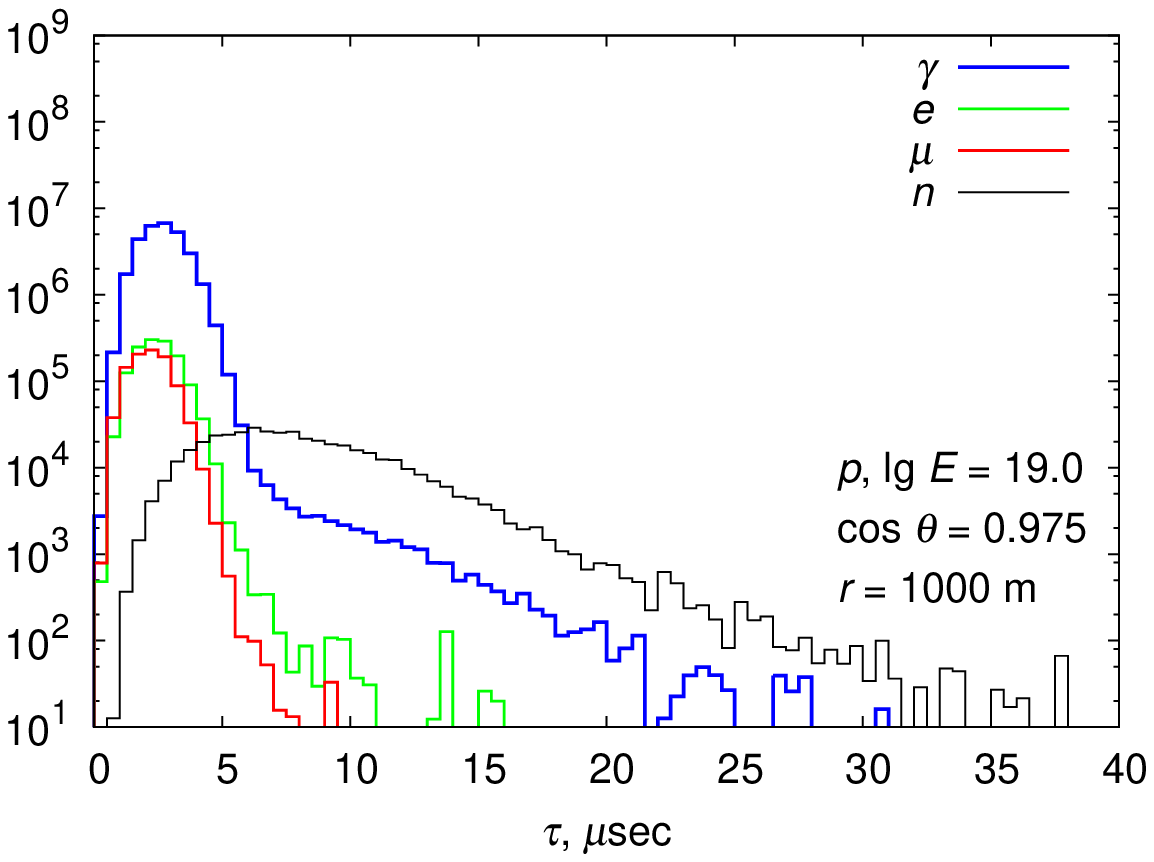}
  \caption{Arrival time distribution of electromagnetic, muon and neutron components of EAS.}
  \label{fig4}
\end{figure}

It is seen from the Fig.~\ref{fig3} that LDF of neutrons is less steep compared to other components, i.e. neutrons can propagate to larger distances from shower axis. At distances further than $2.5$~km from the axis the neutron component exceeds electron and muon components. Hence, neutrons can be detected by products of their moderation in the medium. As a rule, far below the shower maximum and at large distances from the axis it must be slow neutrons of lower energies and products of their moderation can be low-energy electrons. This hypothesis was confirmed when electrons were registered in scintillation detectors with the threshold $\epsilon_{\text{thr.}} \ge 1.8$~eV. Arrival time distributions of electromagnetic, muon and neutron components calculated for primary proton with $E_{0} = 10^{18}$~eV at core distance $r \le 1000$~m is shown on Fig.~\ref{fig4}. From the Fig.~\ref{fig2} it follows that at the distance $1000$~m and with delays by more than $5~\mu$s neutrons become a dominant component of EAS.

\begin{figure}
  \centering
  \includegraphics[width=0.75\textwidth,clip]{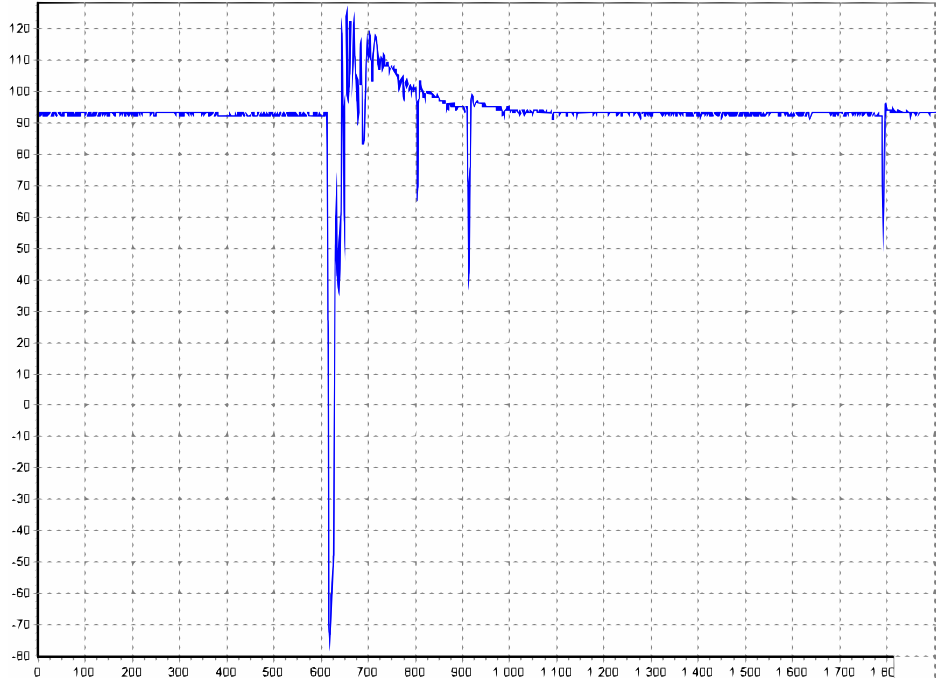}
  \caption{The shower registered on 23-05-2008. $\theta=45.7^{\circ}$, $E_{0} = 4.78 \times 10^{19}$~eV, $r_{\text{obs.lev.}} = 464$~m, $\tau_1 = 1.8~\mu$s, $\tau_2 = 11.8~\mu$s. $s = 1$~m$^2$, $\epsilon_{\text{thr.}} = 1.8$~MeV. The relation of amplitudes from muon and charged components $A_{\mu} / A_s = 0.26$. The division value on horizontal axis is $1~\mu$s.}
  \label{fig5}
\end{figure}

\begin{figure}
  \centering
  \includegraphics[width=0.75\textwidth,clip]{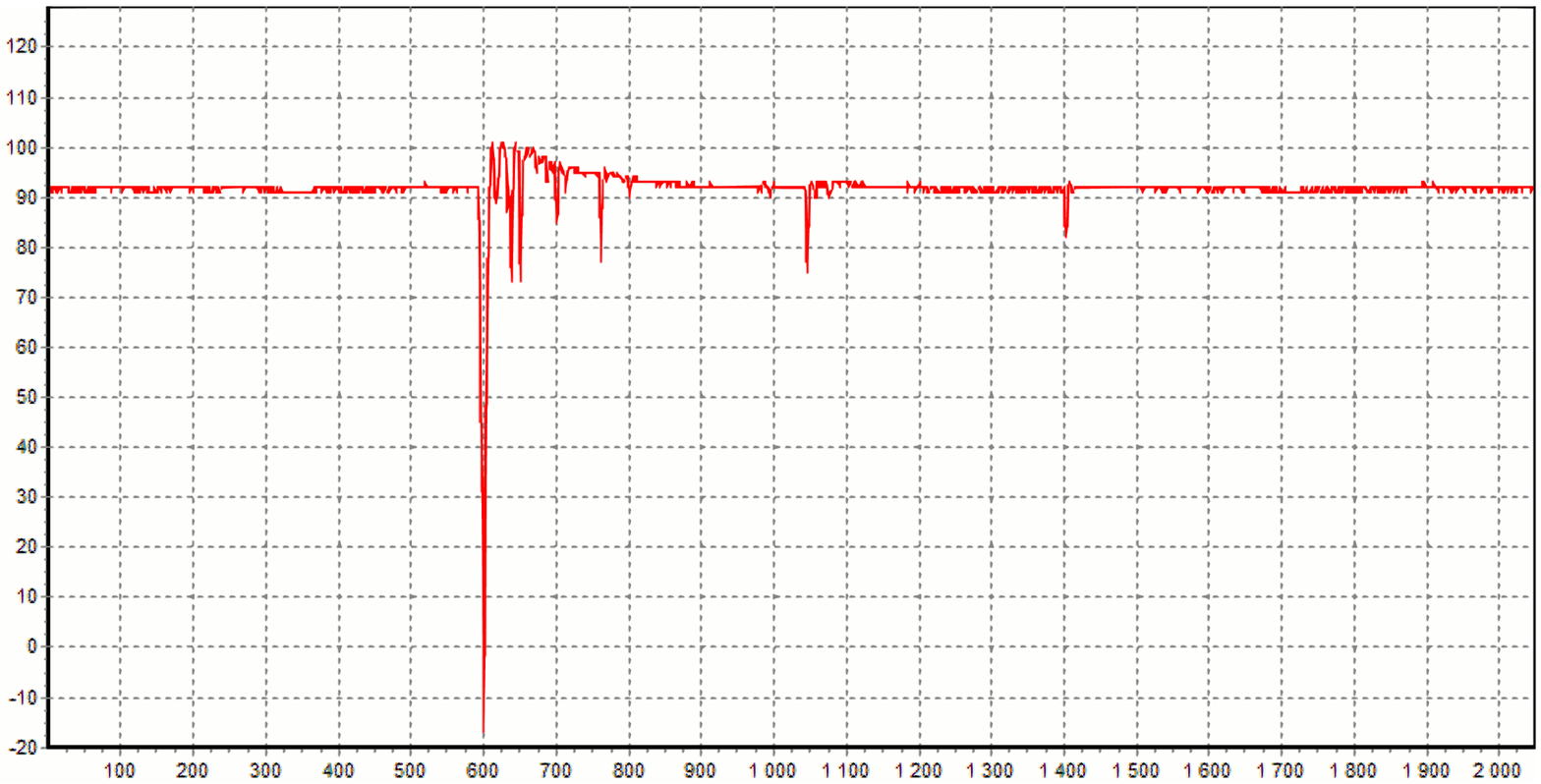}
  \caption{The shower registered on 22-03-2013. $\theta=45.8^{\circ}$, $E_{0} = 1.78 \times 10^{19}$~eV, $r_{\text{obs.lev.}} = 444$~m, $\tau_1 = 4.5~\mu$s, $\tau_2 = 8~\mu$s. $s = 1$~m$^2$, $\epsilon_{\text{thr.}} = 1.8$~MeV. The relation of amplitudes from muon and charged components $A_{\mu} / A_s = 0.22$. The division value on horizontal axis is $1~\mu$s.}
  \label{fig6}
\end{figure}

\begin{figure}
  \centering
  \includegraphics[width=0.49\textwidth,clip]{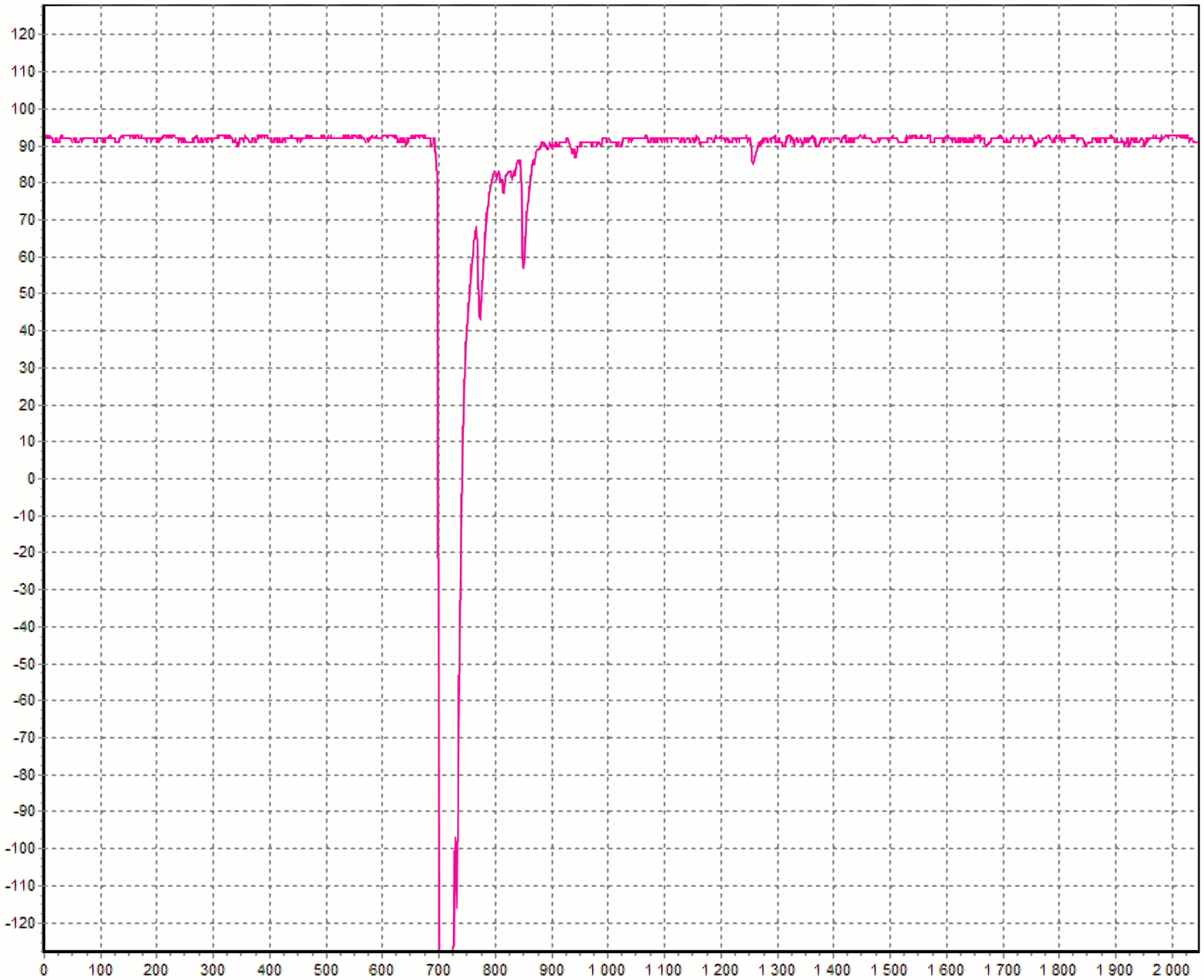}
  \caption{Readings from the detector at $r_{\text{obs.lev.}} = 105$~m in the shower from Fig.~\ref{fig6}. $s = 0.25$~m$^2$, $\epsilon_{\text{thr.}} = 10.5$~MeV. The division value on horizontal axis is $1~\mu$s.}
  \label{fig7}
\end{figure}

\section{The nature of delayed pulses recorded by scintillation detectors}

It is seen from Fig.~\ref{fig5}-\ref{fig7} that particles delayed by $5-7~\mu$s and $11.8~\mu$s were recorded in showers. The pulse with maximal delay was registered by detectors with the area $s = 1$~m$^2$ and threshold $\epsilon_{\text{thr.}} \ge 1.8$~MeV. The amplitude of this pulse exceeded the background noise significantly. It is worth noting that in other detectors including the one with $s = 2$~m$^2$ the signal is extremely weak or is absent completely. It is also should be mentioned that signals delayed by significant time were recorded not in every shower with $E_{0} \ge 10^{19}$~eV. Our task was to collect as many such showers as possible and to analyze them and clarify the nature of the delayed particles.

\begin{table*}
  \centering
  \caption{Characteristics of individual showers with $E_{0} \ge 10^{19}$~eV registered by Yakutsk array compared to predictions by QGSJET01d model. Here $\tau_{1/2}$~--- half-width of a pulse, $r$~--- axis distance, $\rho_{s}, \rho_{\mu}$~--- the density of charged particles and muons correspondingly, $\eta_{300-600}$~--- the slope of the charged particles LDF at $300-600$~m, $\rho_{\mu} / \rho_s$~--- muon fraction with $\epsilon \ge 1$~GeV at $r = 600$~m from the axis. The last column lists the muon fraction according to QGSJET01d model for protons and iron nuclei.}
  \begin{tabular}{lllllllllllll}
    \hline
    date, & time & $\lg{E}$ & $\tau_{1/2}$, & $\theta^{\circ}$ & $\phi^{\circ}$ & $r$, &
    $\rho_{s}$, & $\rho_{\mu}$, & $\eta_{300-600}$ & $\rho_{\mu} / \rho_{s}$, &
    $\rho_{\mu} / \rho_{s}$ & $\rho_{\mu} / \rho_{s}$ \\
dd-mm-yy & & & ns & & & m & 1/m$^2$ & 1/m$^2$ & & (exp.) & (p) & (Fe) \\
    \hline
    11-05-07 & 06:23:31 & 19.35 & 270 & 9.9  & 109.3 & 1298 &  5.3  & 0.72 & -3.32 & 0.14 & 0.15 & 0.20 \\
    15-05-07 & 09:34:33 & 19.29 & 188 & 59.6 & 282.3 & 1000 &  2.6  & 1.44 & -2.35 & 0.55 & --   & --   \\
    19-10-07 & 02:04:13 & 19.33 & 190 & 44.1 &  29.8 &  992 &  4.7  & --   & -2.75 & --   & 0.33 & 0.41 \\
    25-01-08 & 20:34:34 & 19.00 & 202 & 36.0 & 137.9 &  768 &  7.52 & 1.15 & -3.01 & 0.32 & --   & --   \\
    01-05-08 & 19:45:14 & 19.32 & --  & 45.2 & 228.1 & 1298 & 21.3  & 6.30 & -2.75 & 0.30 & --   & --   \\
    01-05-08 & 07:00:34 & 19.17 & --  & 11.2 & 258.2 & 1600 & 1.93  & 0.48 & -3.28 & 0.25 & --   & --   \\
    02-05-08 & 13:21:08 & 19.23 & --  & 19.9 & 330.2 & 1150 & 3.83  & 1.00 & -3.18 & 0.21 & --   & --   \\
    02-01-08 & 08:00:24 & 19.41 & --  & 30.0 & 1.00  & 1400 & --    & --   & -3.12 & 0.26 & --   & --   \\
    08-05-08 & 20:36:02 & 19.18 & --  & 19.8 & 110.1 & 1325 & 2.23  & 0.73 & -3.23 & 0.19 & --   & --   \\
    16-05-07 & 04:16:04 & 19.18 & 170 & 23.8 & 183.6 & 2310 & --    & --   & -3.61 & --   & --   & --   \\
    23-05-08 & 00:58:58 & 19.68 & 165 & 45.7 & 315.5 & 464  & 96.5  & 25.1 & -2.79 & 0.26 & --   & --   \\
    03-01-09 & 03:49:59 & 19.44 & 213 & 42.4 & 14.6  & 944  & 6.34  & 1.63 & -2.88 & 0.26 & --   & --   \\
    22-01-09 & 19:51:52 & 19.64 & 160 & 34.5 & 269.3 & 1112 & 10.2  & 1.77 & -2.88 & 0.17 & --   & --   \\
    22-01-09 & 08:40:35 & 19.56 & 180 & 42.7 & 181.9 & 1167 & 0.80  & 3.47 & -2.89 & 0.23 & --   & --   \\
    21-01-09 & 06:15:14 & 19.10 & 200 & 16.9 & 185.4 & 993  & 4.30  & 0.44 & -3.28 & 0.10 & --   & --   \\
    03-02-09 & 22:35:17 & 19.82 &  90 & 41.2 & 67.5  & 1169 & 11.6  & 1.99 & -3.28 & 0.17 & --   & --   \\
    22-02-09 & 14:08:40 & 19.17 & 230 & 28.5 & 297.4 & 996  & 4.40  & 1.30 & -3.13 & 0.30 & --   & --   \\
    25-02-09 & 22:35:34 & 19.04 & 150 & 7.7  & 122.6 & 1172 & 2.12  & 0.60 & -3.33 & 0.28 & --   & --   \\
    22-03-09 & 16:21:47 & 19.06 & 150 & 32.4 & 84.1  & 549  & 16.7  & 5.30 & -3.02 & 0.32 & --   & --   \\
    10-05-09 & 06:14:18 & 19.03 & 90  & 4.4  & 219.4 & 1383 & 1.14  & 0.52 & -3.55 & 0.46 & --   & --   \\
    22-03-13 & 16:36:25 & 19.25 & --  & 45.8 & 4.3   & 444  & 16.2  & 3.53 & --    & 0.22 & --   & --   \\
    \hline
  \end{tabular}
  \label{tab1}
\end{table*}

After the analysis of the selected showers it was discovered, that pulses delayed by significant time are effectively registered in showers with the energy above $10^{19}$~eV and with the zenith angle $\theta \sim 45^{\circ}$. Such showers are listed in the Table~\ref{tab1}. We suppose that under these conditions the symmetry of shower development is disrupted, unlike in vertical EAS events. In inclined showers, starting from $\theta \ge 45^{\circ}$ a significant separation of particles by their type occurs. The particle composition in over-axial and under-axial parts shifts towards muon component with some content of nucleon component which is transferred to far periphery of particle lateral distribution. As it was pointed in the work~\cite{bib3}, these particles could be low-energy neutrons which, being neutral, don't interact with the material of detector or its container but produce low-energy electron and gamma-quanta when are moderated. These products can be effectively registered by scintillation detectors with $1.8$~MeV threshold, but detectors with $10.5$~MeV don't register them. In case of Yakutsk array, due to permafrost and large amount of snow in winter, the effectiveness of this process may be high enough: snow and permafrost amplifies the moderation of low-energy neutrons leading to production of albedo electrons which produce response in low-threshold detectors. It is clearly seen on Fig.~\ref{fig5}-\ref{fig7}.

\section{Conclusion}

From the analysis of showers with energy above $10^{19}$~eV and different zenith angles, the following conclusions can be summarized:

\begin{itemize}
  \item[a)]
    in vertical showers at the observation level and at moderate distances from the shower axis, signals in scintillation detectors are generated by electromagnetic and muon components, the integration of signals from various sources lasts over $2-3~\mu$s;

  \item[b)]
    single pulses are registered in strongly inclined showers; they are formed strictly by muon component;

  \item[c)]
    when a shower with $\theta \sim 45^{\circ}$ is registered, in the time-base recorded at $r = \sim 400$~m there are pulses delayed by $\tau \ge 5~\mu$s (see Table~\ref{tab1}). According to simulation, the delayed pulses are probably associated with neutron component of EAS, which produces low-energy electrons studying moderation in the material of a detector and surrounding matter. This low energy electrons are registered by scintillation detectors with low thresholds;

  \item[d)]
    as a consequence, there is a possibility of registering pulses from wandering neutrons not associated with EAS by widely spaced detectors (by more than $1000$~m) and, thus, the increased probability of registering a false shower.
\end{itemize}

\acknowledgments
    This work is supported by The Russian Foundation for Basic Research (grant project 12-02-31442\,mol\_a) and by Ministry of Education and Science of the Russian Federation (contract 8404).

\end{document}